\documentclass[12pt]{article}
\usepackage{latexsym}
\usepackage{amssymb}
\usepackage{amsthm}
\usepackage{epsfig}
\usepackage{graphicx}
\graphicspath{{./}{Figures/}}
\usepackage{mathtools} 
\usepackage{enumitem}
\usepackage{xcolor}
\usepackage{easyReview}
\usepackage{algorithm}
\usepackage[noend]{algpseudocode}

\usepackage{hyperref}
\hypersetup{
    linkcolor=blue,  
    linkbordercolor=red,
    linktoc=all,     
    linktocpage
}

\def\dsp{\def\baselinestretch{1.25}\large\normalsize}
\dsp
\allowdisplaybreaks

\begin{document}
\title{Asymptotic Effects of Incident Angle and Lateral Conduction in 
Electromagnetic Skin Heating}
\author{Ulises Jaime-Yepez \; and \;  
Hongyun Wang\footnote{Corresponding author, hongwang@ucsc.edu} 
\\
Department of Applied Mathematics \\
University of California, Santa Cruz, CA 95064, USA\\
\\
Shannon E. Foley \\
U.S. Department of Defense \\
Joint Intermediate Force Capabilities Office \\
Quantico, VA 22134, USA\\
\\
Hong Zhou \\ Department of Applied Mathematics\\
Naval Postgraduate School, Monterey, CA 93943, USA
}

\maketitle

\newpage
\begin{abstract}
Previously we derived the leading term asymptotic solution of temperature distribution 
in skin heating by an electromagnetic beam at an arbitrary incident angle. 
The asymptotic analysis is based on that the penetration depth of the beam into skin is 
much smaller than the size of beam cross-section . 
It allows arbitrary incident angle. 
We expand the temperature in powers of the small depth to lateral scale ratio. 
The incident angle affects all terms in the expansion while the lateral heat conduction 
appears only in terms of positive even powers. The previously obtained leading term solution 
captures only the main effect of incident angle. 
The main effect of lateral heat conduction is contained in the second order term, which 
is mathematically negligible in the limit of small depth to lateral scale ratio. 
At a moderate length scale ratio (e.g., 0.1), however, the contribution from lateral conduction 
is quite significant and needs to be included in a meaningful approximate solution. 
In this study, we derive closed form analytical expressions for the first order and the second 
order terms in the asymptotic expansion. 
The resulting asymptotic solution is capable of predicting the temperature distribution 
accurately including the effects of both incident angle and lateral heat conduction 
even at a moderate length scale ratio. 
\end{abstract}


\clearpage
\renewcommand*\contentsname{Table of contents}
\tableofcontents
\clearpage

\section{Introduction}
Millimeter-wave (MMW) systems operate using electromagnetic waves
of wavelength in millimeters corresponding to the frequency range of 30-300 GHz. 
These systems are present in many civilian and defense applications. 
The main biological effects of MMW exposure on human are attributed to the thermal 
effects in heating skin tissue, which have been studied extensively. 
A number of these studies were compiled in a special issue of the Journal of Directed Energy \cite{Whitmore_2021, Miller_2021, Cook_2021, Cook_2021B, Haeuser_2021, 
Cobb_2021, Parker_2021, Parker_2021B}. 
One particular thermal effect on skin is the possible thermal burn damage \cite{Parker_2024}. 

The primary thermal effect of MMW exposures is skin temperature increase 
driven by the electromagnetic power absorbed into skin. 
The penetration depth of an electromagnetic wave into skin is the depth at which 
$(1/e)$ fraction of the electromagnetic power survives and the rest has been absorbed. 
For frequencies between 30 and 300 GHz, the penetration depth is well below millimeter
\cite{Parker_2017}\cite{Parker_2017B}. 
Consequently, the electromagnetic power is practically absorbed within 
one millimeter of skin depth. This rapid absorption of electromagnetic power in 
a thin skin layer creates a large temperature gradient in the depth direction. 
In the lateral directions along skin surface, the heating varies with the power density 
passing through the skin surface over the beam cross section, which typically spans centimeters.
This separation of length scales in the depth and in lateral directions gives 
a small ratio of depth to lateral scale $\varepsilon$, which is the foundation for 
asymptotic analysis.  

The effect of heat conduction is inversely proportional to the square of the length scale. 
When the depth to lateral scale ratio $\varepsilon$ is small, 
relative to the dominant conduction in depth, the effect of conduction in lateral directions 
is of the order $\varepsilon^2$. 
For small $\varepsilon$, we expand the skin temperature distribution in powers of $\varepsilon$. 
The main effect of lateral heat conduction is in the second order ($\varepsilon^2$) term. 
The leading ($\varepsilon^0$) and the first order ($\varepsilon$) terms are independent 
of lateral heat conduction. 

The effect of incident angle on the electromagnetic heating is more complicated. It percolates 
into all powers of $\varepsilon$. 
The intrinsic beam description specifies beam size and power distribution over a cross-section 
perpendicular to the beam axis. First, given the intrinsic beam description, 
the incident angle affects the beam spot projected onto 
the skin surface. At a non-trivial incident angle, the projected beam spot is elongated 
and accordingly the beam power density over the projected beam spot is diluted. 
The projected power density over the skin surface affects the amplitude of the heating source. 
Second, inside the skin tissue, the beam propagates and is absorbed along the refracted 
direction, not the depth direction perpendicular to the skin surface. 
The refracted direction increases the beam's propagation distance per depth, and thus, 
increases the absorption per depth. 
In our mathematical formulation, the increased absorption rate corresponds to 
a stretching of the depth coordinate in the function of heating source vs depth. 
Both factor 1 (projected power density on the skin surface) and factor 2 
(stretching of the depth coordinate in the heating source) are attributed solely to the incident angle,
independent of the depth to lateral scale ratio $\varepsilon$. 
In the non-dimensionalized mathematical model, if we hypothetically set $\varepsilon = 0$, 
both factors 1 and 2 are still present. 
Third, the propagation along the refracted direction shifts the heating source at each depth level 
by a depth dependent amount in a lateral direction. 
In the non-dimensionalized mathematical model, this depth dependent lateral shift is 
tangled with the depth to lateral scale ratio $\varepsilon$. 
In the hypothetical situation of $\varepsilon = 0$, the effect of the lateral shift disappears. 
For $\varepsilon = O(1)$, the 3D heat equation needs to be solved numerically
due to the $O(1)$ presence of the depth-dependent lateral shifting in heating. 
At small $\varepsilon > 0$, an asymptotic approximation of the temperature distribution is 
obtained by expanding in powers of $\varepsilon$. 
Previously, we derived the leading term of the asymptotic expansion, which 
contains the main effect of only factors 1 and 2 above \cite{Wang_2025}. 
In this study, we derive a second order asymptotic solution containing up to $O(\varepsilon^2)$
terms in the expansion. This high order asymptotic solution captures i) the effects 
of all factors attributed to the incident angle and ii) the effect of lateral heat conduction. 
One possible utility of a fast and viable temperature solver is in the inverse problem of 
inferring internal temperature from a time series of measured surface temperatures 
\cite{WBZ_2023}.

This paper is organized as follows. 
Section 2 reviews the non-dimensional mathematical model developed in \cite{Wang_2025}
for skin heating by an incident electromagnetic beam. 
The model includes the full effect of incident angle and lateral heat conduction. 
So far, a closed form exact solution to this full model is not yet known. 
We solve it asymptotically by expanding the skin temperature distribution
in powers of the small depth-to-lateral scale ratio $\varepsilon$. 
In Section 3, we construct initial boundary value problems (IBVPs) governing individual 
terms in the asymptotic expansion. Then we solve these IBVPs to derive closed form 
analytical expressions for the $O(\varepsilon)$ and $O(\varepsilon^2)$ terms in the 
expansion. 
In Section 4, we dissect the effects of incident angle and lateral heat conduction, 
and examine numerically the performance of these asymptotic solutions obtained. 
In that task, we use the central finite difference method to solve the full 3D heat equation 
on a fine grid and use the accurate numerical solution as the exact solution when 
evaluating errors in asymptotic solutions. 
Finally, Section 5 summarizes the main results of the study. 

\section{Governing Model of Electromagnetic Heating in Skin} \label{temp_evolu}
\subsection{Skin, Incident Beam and Two Coordinate Systems}
\begin{figure}[!h]
\vskip 0.4cm
\begin{center}
\psfig{figure=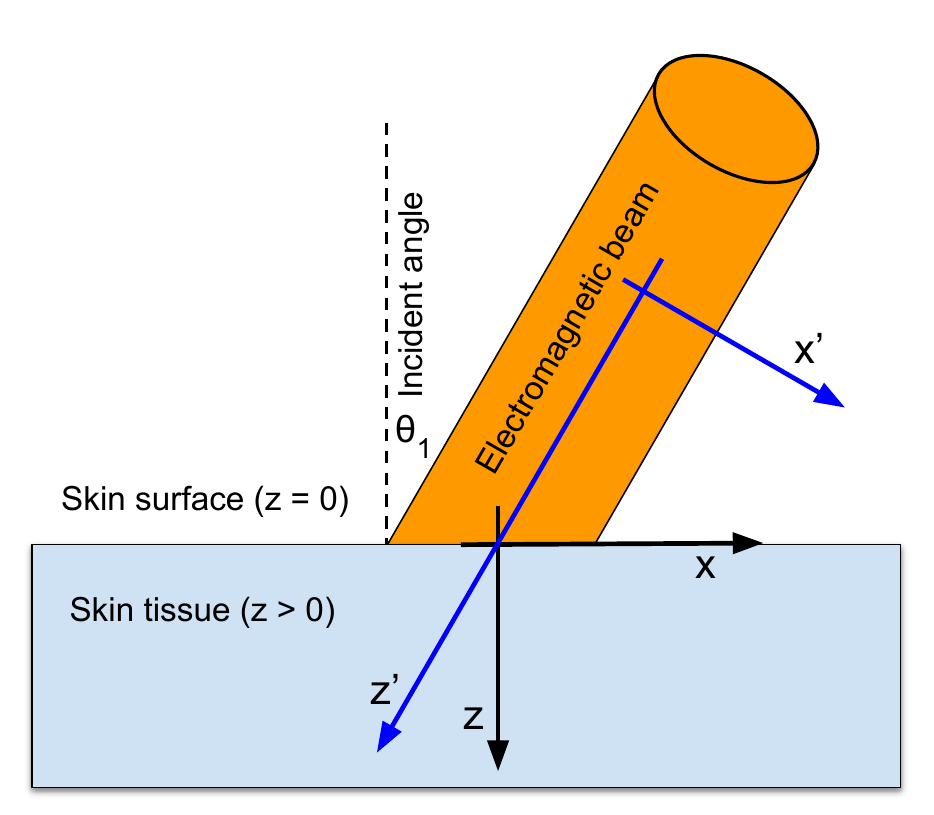, height=3.5in} 
\end{center}
\vskip -0.8cm
\caption{Schematics of the skin, the incident beam, the skin coordinate system
$(x, y, z)$, and the beam intrinsic coordinate system $(x', y', z')$.} 
\label{fig_01}
\end{figure}

We consider the situation where a skin tissue with a flat surface is exposed to 
an incident electromagnetic wave, not necessarily perpendicular to the skin surface. 
We establish the coordinate system as described in our previous study \cite{Wang_2025}. 
In Figure \ref{fig_01}, the normal direction into the skin tissue is defined as 
 the positive $z$-axis, with $z=0$ at the skin surface. 
 The $z$-coordinate represents the depth into the skin. 
The incident beam is tilted away from the $z$-axis by an angle $\theta_1$, 
referred to as the incident angle. 
We select the $x$-axis such that the tilting of the incident beam is in the $(x, z)$-plane and 
toward the $x$-axis. 
Once the $z$- and $x$-axes are defined, the $y$-axis is automatically determined 
from the right hand rule. 

The exposure setup in Figure \ref{fig_01} is specified by 
i) the intrinsic description of the electromagnetic beam with respect to its intrinsic axis, 
independent of the presence of skin, and ii) the incident angle of the beam relative 
to the skin surface. 
Let $(x', y', z')$ denote the coordinates in the intrinsic frame attached to the beam 
where the $z'$-axis is along the beam axis of propagation, the $y'$-axis coincides with the 
$y$-axis, and the $x'$-axis is tilted away from the $x$-axis toward negative $z$ 
in the $(x, z)$-plane, as illustrated in Figure \ref{fig_01}. 

Let $P_d^{\text{(intr)}}(x', y') = P_d^{(i)}q(x', y')$ be the power density over a beam cross-section 
perpendicular to the axis of propagation where $P_d^{(i)}$ is the beam's intrinsic power density 
at the center, and $q(x', y')$ describes the relative distribution satisfying $q(0, 0) = 1$. 
Power density $P_d^{\text{(intr)}}(x', y')$ is intrinsic to the beam, 
independent of the incident angle, $\theta_1$. 
The projected beam spot onto the skin surface 
is elongated by a factor of $1/\cos\theta_1$ in the $x$-direction. 
Accordingly the power density projected onto the $(x, y)$-plane, denoted by 
$P_d^{\text{(proj)}}(x, y)$, is reduced by a factor of $\cos\theta_1$.  
\begin{align*}
& P_d^{\text{(proj)}}(x, y) 
 =  \cos\theta_1 P_d^{(i)} q(x \cos\theta_1, y) 
\end{align*}

\subsection{Absorption of Electromagnetic Wave in Skin}
At the skin surface, a fraction $\alpha$ of the arriving power density passing 
into the skin. 
\begin{equation}
P_d^{\text{(pass)}}(x, y)  = \underbrace{\alpha \cos\theta_1 P_d^{(i)}}_{
\equiv P_d^{(a)}} \; \underbrace{q(x \cos\theta_1, y)}_{\equiv f(x, y)}
\equiv P_d^{(a)} f(x, y) 
\label{Pda_fxy_def}
\end{equation}
where $P_d^{(a)} \equiv \alpha \cos\theta_1 P_d^{(i)}$ is the power density at the beam center
passing into the skin, which is absorbed inside skin. $f(x, y) \equiv q(x \cos\theta_1, y)$ 
is the relative distribution of beam power density projected on the skin surface, 
satisfying $f(0, 0) = 1$. 

\begin{figure}[!h]
\vskip 0.5cm
\begin{center}
\psfig{figure=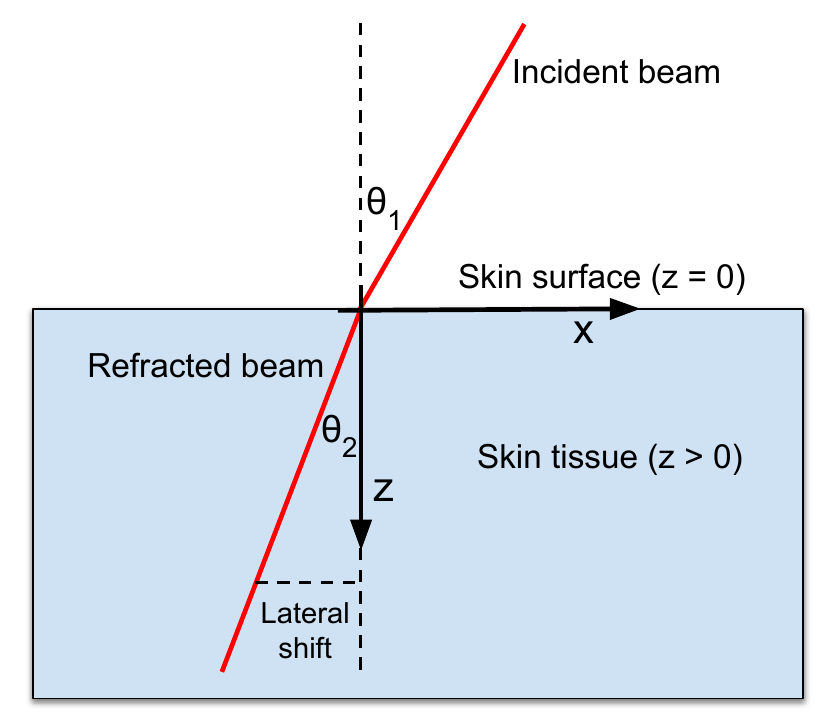, width=3.5in} 
\end{center}
\vskip -0.6cm
\caption{The refracted direction describes the propagation distance of beam inside skin vs depth
and the lateral shift of the high power density region vs depth.}
\label{fig_02}
\end{figure}

Inside skin, the electromagnetic wave propagates along the refracted direction, 
which is at angle $\theta_2$ to the $z$-axis. 
The refracted angle $\theta_2 $ is governed by Snell's law \cite{Bezugla_2024}:
\[ \sin\theta_2 = \sin\theta_1 \frac{n_\text{air}}{n_\text{skin}} \] 
Note that $P_d^{\text{(pass)}}(x, y)$ is the power per unit area of the skin surface 
($z = 0$), which is perpendicular to neither the incident direction nor the refracted direction. 
The electromagnetic thermal heating in skin is caused by the beam power absorbed 
per unit volume. To calculate the heat source, we study the decay of power density 
at each level of depth ($z$). Based on Beer-Lambert Law, the decay is exponential
with respect to the propagation distance. 
For the refracted beam, the propagation distance is related to the depth by the refraction angle.  
\[ \text{propagation distance} = \frac{z}{\cos\theta_2} \] 
As the refracted beam propagates inside skin, it shifts the active heating area toward 
the negative $x$-direction as illustrated in Figure \ref{fig_02}. The amount of lateral shift is 
depth dependent. 
The power density at $(x, y)$ at depth $z$ is the surviving part of the power density 
passing through $(x_s, y_s)$ on the skin surface. Tracing from $(x, y, z)$ at depth $z$
along the refracted direction to $(x_s, y_s, 0)$ on the skin surface, we have 
\[ x_s = x+z \tan\theta_2, \qquad y_s = y \]
We combine the effects of exponential decay (absorption) and lateral shift to calculate 
$P_d^{\text{(pass)}}(x, y; z) $, the power per unit area at depth level $z$. 
\[ P_d^{\text{(pass)}}(x, y; z)  =  P_d^{(a)} 
\underbrace{f(x_s, y)}_{\text{lateral shift}} 
\underbrace{e^{\frac{-\mu}{\cos\theta_2}z}}_{\text{exponential decay}}, \qquad
x_s = x+z \tan\theta_2 \] 
where $\mu$ is the power absorption coefficient of skin for the electromagnetic frequency used.
Let $S(x, y, z)$ denote the rate of heat from the absorbed electromagnetic 
power per unit volume at location $(x, y, z)$. 
Conservation of energy gives 
\begin{align*}
& S(x, y, z) = -\frac{d}{dz}\Big(P_d^{\text{(pass)}}(x, y; z)\Big|_{x_s, y \text{ fixed}}\Big) 
= P_d^{(a)} f(x_s, y) \frac{\mu}{\cos\theta_2} 
e^{\frac{-\mu}{\cos\theta_2}z} \\[1ex]
& \qquad \qquad = P_d^{(a)} f(x+z \tan\theta_2, y) \frac{\mu}{\cos\theta_2} 
e^{\frac{-\mu}{\cos\theta_2}z}
\end{align*}
In the above, fixing $(x_s, y)$ corresponds to that power flows along the direction of propagation. 
$S(x, y, z)$ serves as the heat source in the skin temperature evolution. 

\subsection{Skin Temperature Evolution}
To model the temperature evolution, we proceed with the following assumptions. 
\begin{enumerate}
    \item The skin's material properties are uniform in space, independent of $(x, y, z)$.
    \item The skin temperature prior to electromagnetic exposure is uniform in space, which  
    is called the baseline skin temperature \cite{Walters_2000} and is denoted by $T_\text{base}$. 
    \item The electromagnetic power input per unit area at the skin surface is 
    much larger than the rate of heat loss per unit area at the skin surface. 
    Thus, in the model, we neglect the heat loss at the skin surface during the short 
    time of exposure. 
     \item The thickness of skin tissue is much larger than both the electromagnetic 
     penetration depth and the thermal diffusion depth over the short time of exposure. 
     Mathematically, we treat the skin as semi-infinite 
     extending to $z = +\infty$. 
\end{enumerate}
Let $T(x, y, z, t)$ denote the skin temperature at position $(x, y, z)$ at time $t$. 
The governing equation for $T(x, y, z, t)$ is derived from conservation of energy. 
The initial and boundary conditions for $T(x, y, z, t)$ follow directly from Assumptions 2 and 3.
\begin{equation}
\begin{dcases}
\underbrace{\rho_m C_p\frac{\partial T}{\partial t}}_{\substack{
\text{rate of change}\\ \text{of heat in skin}}}
= \underbrace{k \Big(\frac{\partial^2T}{\partial x^2}
+\frac{\partial^2T}{\partial y^2}+\frac{\partial^2T}{\partial z^2}
\Big)}_{\substack{
\text{heat flux from}\\ \text{conduction}}} 
+ \underbrace{P_d^{(a)} f(x+z \tan\theta_2, y) \frac{\mu}{\cos\theta_2} 
e^{-\frac{\mu}{\cos\theta_2} z}}_{
\substack{\text{heat source from}\\ \text{absorbed power}}} \\[2ex]
\underbrace{\frac{\partial T(x, y, z, t)}{\partial z} \Big|_{z=0} = 0}_{
\text{no heat loss at skin surface}}  \\[1ex]
\underbrace{T(x, y, z, 0) = T_\text{base}}_{\text{baseline temperature}}
\end{dcases}
\label{HEq_1}
\end{equation}
In \eqref{HEq_1}, $\rho_m $ denotes the mass density, $C_p$ the specific heat capacity, 
and $k$ the heat conductivity of the skin. 
In addition to the explicit presence of refracted angle $\theta_2$ in \eqref{HEq_1}, the effect of 
incident angle is also implicitly in $P_d^{(a)} \equiv (\alpha \cos\theta_1  P_d^{(i)})$ 
and in $f(x, y) \equiv q(x \cos\theta_1, y)$. 
Here $P_d^{(a)}$ is the power density passing into the skin at the beam center 
and $f(x, y)$ is the relative distribution of power density projected on the skin surface. 

\subsection{Nondimensionalization}
We list the scales for various physical quantities and the resulting 
nondimensional quantities. Here we use the simple notation $X$ without the subscript
$_\text{nondim}$ for all nondimensional quantities. 
For clarity, the original physical quantities are denoted by $X_\text{phy}$
\cite{Wang_2020}. 
\begin{equation}
\begin{dcases}
 z_s \equiv \frac{1}{\mu}, \qquad z \equiv \frac{z_\text{phy}}{z_s} \\[0ex]
r_s \equiv (\text{lateral length scale}), \qquad 
(x, y) \equiv \frac{(x_\text{phy}, y_\text{phy})}{r_s} \\[1ex]
t_s \equiv \frac{\rho_m C_p}{k \mu^2},\qquad t \equiv \frac{t_\text{phy}}{t_s} \\[1ex]
T_s\equiv (\text{temperature scale}), \\ 
\qquad T(x, y, z, t) \equiv \frac{T_\text{phy}(x_\text{phy}, y_\text{phy}, z_\text{phy}, t_\text{phy})
-T_\text{base,phy}}{T_s} \\[1ex]
P_s\equiv k \mu T_s, \quad P_\text{d}^{(i)} \equiv\frac{P_\text{d,phy}^{(i)}}{P_s}, 
\quad P_d^{(a)} \equiv\frac{P_\text{d,phy}^{(a)}}{P_s} = (\alpha \cos\theta_1  P_d^{(i)}) \\[1ex] 
f(x, y) \equiv f_\text{phy}(x_\text{phy}, y_\text{phy}) 
\end{dcases}
\label{scales_nd_qs} 
\end{equation}
In the above, the lateral length scale $r_s$ is usually set to the beam size; 
$T_\text{base,phy}$ is the physical baseline temperature; 
the temperature scale $T_s$ is usually set to the physical temperature 
increase from the baseline to the activation of skin thermal nociceptors or is set 
to any physically relevant temperature increase. 
The physical temperature as a function of physical variables 
$T_\text{phy}(x_\text{phy}, y_\text{phy}, z_\text{phy}, t_\text{phy})$ 
is governed by the initial boundary value problem (IBVP) \eqref{HEq_1}. 
The nondimensional version of \eqref{HEq_1} for $T(x,y,z,t)$ is 
\begin{equation}
\begin{dcases}
\displaystyle \frac{\partial T}{\partial t} 
= \varepsilon^2 \Big(\frac{\partial^2T}{\partial x^2}
+\frac{\partial^2T}{\partial y^2}\Big)+\frac{\partial^2T}{\partial z^2}
+P_d^{(a)} f(x+\varepsilon z \tan\theta_2, y) 
\frac{1}{\lambda} e^{-z/\lambda}, 
\quad \lambda \equiv \cos\theta_2 \\[2ex]
\displaystyle \frac{\partial T(x, y, z, t)}{\partial z} \bigg|_{z=0}=0 \\[1ex]
\displaystyle T(x, y, z, 0) = 0 
\end{dcases}
\label{IBVP_1}
\end{equation}
where $\displaystyle \varepsilon \equiv \frac{z_s}{r_s} $ is 
the depth-to-lateral scale ratio. IBVP \eqref{IBVP_1} governs the evolution 
of nondimensional temperature $T(x,y,z,t)$. It includes the full effect of incident angle 
and lateral heat conduction, and is valid for arbitrary $\theta_1$, $\theta_2$ and 
$\varepsilon$. In the next section, we construct asymptotic solutions of \eqref{IBVP_1}
in the case of small depth-to-lateral scale ratio $\varepsilon$. 
These asymptotic approximations are valid for arbitrary $\theta_1$ and $\theta_2$
as long as $\varepsilon$ is small. 
%

\section{Asymptotic Solutions of 3D Skin Temperature} \label{asy_sol}
We construct asymptotic solutions of \eqref{IBVP_1} in the case of  
$\displaystyle \varepsilon \equiv \frac{z_s}{r_s} \ll 1$. 
\subsection{Asymptotic Formulation} \label{asy_formu}
In \eqref{IBVP_1}, the small parameter $\varepsilon$ appears in the lateral heat conduction 
$\varepsilon^2 (\frac{\partial^2 T}{\partial x^2}+\frac{\partial^2 T}{\partial y^2})$ and 
in the heat source $P_d^{(a)} f(x+\varepsilon z \tan\theta_2, y) 
\frac{1}{\lambda} e^{-z/\lambda}$, which contains a depth dependent lateral shift 
caused by a non-trivial incident angle. 
We expand the heat source in powers of $\varepsilon$.
\[ f(x+\varepsilon z \tan\theta_2, y) = f(x, y) + \underbrace{\varepsilon\, f_x(x, y) z \tan\theta_2
+ \varepsilon^2\, \frac{1}{2}f_{xx} (x, y) z^2 \tan^2\theta_2 + \cdots}_{
\text{effect of depth dependent lateral shift}} \]
In \eqref{IBVP_1}, $O(\varepsilon^2 )$ terms come
from two sources: i) the lateral heat conduction and ii) the lateral shift in the heat source. 
It is sensible to distinguish the contributions from these two sources. 
Thus, we adopt the asymptotic form below. 
\begin{align}
& T(x, y, z, t) = T^{(0)}(x, y, z, t) + \varepsilon\, T^{(1)}(x, y, z, t) \nonumber \\[1ex]
& \qquad \qquad + \underbrace{\varepsilon^2\, T^{(2,A)}(x, y, z, t)}_{\text{lateral conduction}} 
+ \varepsilon^2\, T^{(2,B)}(x, y, z, t) + \cdots 
\label{asymp_form}
\end{align}
Here $\varepsilon^2\, T^{(2,A)}$ is attributed to the presence of lateral heat conduction 
$\varepsilon^2 (\frac{\partial^2 T}{\partial x^2}+\frac{\partial^2 T}{\partial y^2})$ in 
\eqref{IBVP_1} while the effect of incident angle appears in all terms since both 
$P_d^{(a)} \equiv (\alpha \cos\theta_1  P_d^{(i)})$ and $f(x, y) \equiv q(x \cos\theta_1, y)$ 
vary with the incident angle. 
Substituting expansion form \eqref{asymp_form} into IBVP \eqref{IBVP_1}, 
we obtain the IBVP for each of $T^{(0)}$, $T^{(1)}$, $T^{(2,A)}$ and $T^{(2,B)}$. 

\subsubsection{IBVP for $T^{(0)}(x, y, z, t)$}
Substituting expansion form \eqref{asymp_form} into IBVP \eqref{IBVP_1}, 
we match the $O(1)$ terms from $T^{(0)}$ and the $O(1)$ term from the heat source
$P_d^{(a)} f(x+\varepsilon z \tan\theta_2, y) \frac{1}{\lambda} e^{-z/\lambda}$.
\begin{equation}
\begin{dcases}
\displaystyle \big(\partial_t - \partial_{zz} \big) T^{(0)} = 
P_d^{(a)} f(x, y) \frac{1}{\lambda} e^{-z/\lambda }, \quad 
\lambda \equiv \cos\theta_2 \\[1ex]
\displaystyle \partial_z T^{(0)}(x, y, z, t) \Big|_{z=0}=0 \\[1ex]
\displaystyle T^{(0)}(x, y, z, 0) = 0 
\end{dcases}
\label{IBVP_T0}
\end{equation}

\subsubsection{IBVP for $T^{(1)}(x, y, z, t)$}
Substituting expansion form \eqref{asymp_form} into IBVP \eqref{IBVP_1}, 
we match the $O(\varepsilon)$ terms from $T^{(1)}$ and the $O(\varepsilon)$ term 
from the heat source
$P_d^{(a)} f(x+\varepsilon z \tan\theta_2, y) \frac{1}{\lambda} e^{-z/\lambda}$.
\begin{equation}
\begin{dcases}
\displaystyle \big(\partial_t - \partial_{zz} \big) T^{(1)} 
= P_d^{(a)} f_x(x, y) \tan{\theta_2}\frac{z}{\lambda} e^{-z/\lambda }, \quad 
\lambda \equiv \cos\theta_2 \\[2ex]
\displaystyle \partial_z T^{(1)}(x, y, z, t) \Big|_{z=0}=0 \\[1ex]
\displaystyle T^{(1)}(x, y, z, 0) = 0 
\end{dcases}
\label{IBVP_T1}
\end{equation}

\subsubsection{IBVP for $T^{(2,A)}(x, y, z, t)$}
Substituting expansion form \eqref{asymp_form} into IBVP \eqref{IBVP_1}, 
we match the $O(\varepsilon^2)$ terms from $T^{(2,A)}$ and the $O(\varepsilon^2)$ term 
from $T^{(0)}$. We leave out the $O(\varepsilon^2)$ terms from $T^{(2,B)}$
and the $O(\varepsilon^2)$ term from 
$P_d^{(a)} f(x+\varepsilon z \tan\theta_2, y) \frac{1}{\lambda} e^{-z/\lambda}$. 
These $O(\varepsilon^2)$ terms will be matched separately. 
\begin{equation}
\begin{dcases}
\displaystyle \big(\partial_t - \partial_{zz} \big) T^{(2,A)}
=  \nabla^2_{(x, y)} T^{(0)} \\[2ex]
\displaystyle \partial_z T^{(2,A)}(x, y, z, t) \Big|_{z=0}=0 \\[1ex]
\displaystyle T^{(2,A)}(x, y, z, 0) = 0 
\end{dcases}
\label{IBVP_T2A}
\end{equation}

\subsubsection{IBVP for $T^{(2,B)}(x, y, z, t)$}
Substituting expansion form \eqref{asymp_form} into IBVP \eqref{IBVP_1}, 
we match the $O(\varepsilon^2)$ terms from $T^{(2,B)}$ and the $O(\varepsilon^2)$ term 
from $P_d^{(a)} f(x+\varepsilon z \tan\theta_2, y) \frac{1}{\lambda} e^{-z/\lambda}$. 
\begin{equation}
\begin{dcases}
\displaystyle \big(\partial_t - \partial_{zz} \big) T^{(2,B)}
= P_d^{(a)} \frac{1}{2}f_{xx}(x, y) \tan^2{\theta_2}\frac{z^2}{\lambda} e^{-z/\lambda }, \quad 
\lambda \equiv \cos\theta_2 \\[2ex]
\displaystyle \partial_z T^{(2,B)}(x, y, z, t) \Big|_{z=0}=0 \\[1ex]
\displaystyle T^{(2,B)}(x, y, z, 0) = 0 
\end{dcases}
\label{IBVP_T2B}
\end{equation}

\subsection{Reduced IBVPs based on Separable Dependence}
Asymptotic terms, $T^{(0)}$, $T^{(1)}$, $T^{(2,A)}$ and $T^{(2,B)}$, in 
\eqref{asymp_form} are functions of $(x, y, z, t)$, governed respectively 
by IBVPs \eqref{IBVP_T0}, \eqref{IBVP_T1}, \eqref{IBVP_T2A}, and \eqref{IBVP_T2B}
obtained above. 
A key feature of these IBVPs is that the solution has a separable dependence on $(z, t)$ and $(x, y)$. 
This key feature greatly facilitates the analytical solution of these IBVPs. 

We now establish the separable dependence in 
IBVPs \eqref{IBVP_T0}, \eqref{IBVP_T1}, \eqref{IBVP_T2A}, and \eqref{IBVP_T2B}. 
All these IBVPs have the same homogeneous initial condition, 
the same homogeneous boundary condition, 
and the common PDE structure of the heat equation with a source term. 
\begin{equation}
\begin{dcases}
\displaystyle \big(\partial_t - \partial_{zz} \big) T =  G(x, y, z, t) \\[1ex]
\displaystyle \partial_z T(x, y, z, t) \Big|_{z=0}=0 \\[1ex]
\displaystyle T(x, y, z, 0) = 0 
\end{dcases}
\label{IBVP_sepa}
\end{equation} 
If source $G$ has a separable dependence: 
$G(x, y, z, t) = h(x, y) F(z, t)$, then the separable dependence carries to solution $T$ with 
the same lateral distribution $h(x, y)$. 
\[ T(x, y, z, t) = h(x, y) W(z, t) \] 
The governing IBVP in $(x, y, z, t)$ for $T$ is reduced in dimensions to an IBVP
in $(z, t)$ for $W$. 
\begin{equation}
\begin{dcases}
\displaystyle \big(\partial_t - \partial_{zz} \big) W(z, t) =  F(z, t) \\[1ex]
\displaystyle \partial_z W(z, t) \Big|_{z=0}=0 \\[1ex]
\displaystyle W(z, 0) = 0 
\end{dcases}
\label{IBVP_zt}
\end{equation} 
In IBVPs \eqref{IBVP_T0}, \eqref{IBVP_T1}, and \eqref{IBVP_T2B}, 
we check separable dependence of the source term. 
\begin{align}
\eqref{IBVP_T0}{:} &\;\; \text{source} = \Big(P_d^{(a)} f(x, y)\Big) 
\frac{1}{\lambda} e^{-z/\lambda } \nonumber \\
& \quad \Longrightarrow \quad T^{(0)}(x, y, z, t) = 
\Big(P_d^{(a)} f(x, y)\Big) W^{(0)}(z, t) \label{T0_sepa_dep} \\[1ex] 
\eqref{IBVP_T1}{:} &\;\; \text{source} = \Big(P_d^{(a)} f_x(x, y) \tan{\theta_2} \Big)
\frac{z}{\lambda} e^{-z/\lambda } \nonumber \\
& \quad \Longrightarrow \quad T^{(1)}(x, y, z, t) = 
\Big(P_d^{(a)} f_x(x, y) \tan{\theta_2} \Big) W^{(1)}(z, t) \label{T1_sepa_dep} \\[1ex] 
\eqref{IBVP_T2B}{:} &\;\; \text{source} = \Big(P_d^{(a)} \frac{1}{2}f_{xx}(x, y) \tan^2{\theta_2} \Big) \frac{z^2}{\lambda} e^{-z/\lambda } \nonumber \\
& \quad \Longrightarrow \quad T^{(2,B)}(x, y, z, t) = 
\Big(P_d^{(a)} \frac{1}{2}f_{xx}(x, y) \tan^2{\theta_2} \Big) W^{(2,B)}(z, t) 
 \label{T2B_sepa_dep}
\end{align}
In \eqref{IBVP_T2A}, with $T^{(0)} = \big(P_d^{(a)} f(x, y) \big) W^{(0)}(z, t) $, the source term is 
\begin{align}
\eqref{IBVP_T2A}{:} &\;\; \text{source} = \nabla^2_{(x, y)} T^{(0)} = 
\Big(P_d^{(a)} \nabla^2_{(x, y)} f(x, y) \Big) W^{(0)}(z, t) \nonumber \\
& \quad \Longrightarrow \quad T^{(2,A)}(x, y, z, t) = 
\Big(P_d^{(a)} \nabla^2_{(x, y)} f(x, y) \Big) W^{(2,A)}(z, t) 
 \label{T2A_sepa_dep}
\end{align} 
Below are the IBVPs for $W^{(0)}$, $W^{(1)}$, $W^{(2,A)}$ and $W^{(2,B)}$.
These IBVPs are in variables $(z, t)$, reduced from those in variables $(x, y, z, t)$ 
for $T^{(0)}$, $T^{(1)}$, $T^{(2,A)}$ and $T^{(2,B)}$. 

\subsubsection{Reduced IBVP for $W^{(0)}(z, t)$ in $T^{(0)}$}
\begin{equation}
\begin{dcases}
\displaystyle \big(\partial_t - \partial_{zz} \big) W^{(0)}(z, t) = 
\frac{1}{\lambda} e^{-z/\lambda }, \quad 
\lambda \equiv \cos\theta_2 \\[1ex]
\displaystyle \partial_z W^{(0)}(z, t) \Big|_{z=0}=0 \\[1ex]
\displaystyle W^{(0)}(z, 0) = 0 
\end{dcases}
\label{IBVP_W0}
\end{equation}

\subsubsection{Reduced IBVP for $W^{(1)}(z, t)$ in $T^{(1)}$}
\begin{equation}
\begin{dcases}
\displaystyle \big(\partial_t - \partial_{zz} \big) W^{(1)}(z, t) = 
\frac{z}{\lambda} e^{-z/\lambda }, \quad 
\lambda \equiv \cos\theta_2 \\[1ex]
\displaystyle \partial_z W^{(1)}(z, t) \Big|_{z=0}=0 \\[1ex]
\displaystyle W^{(1)}(z, 0) = 0 
\end{dcases}
\label{IBVP_W1}
\end{equation}

\subsubsection{Reduced IBVP for $W^{(2,A)}(z, t)$ in $T^{(2,A)}$}
\begin{equation}
\begin{dcases}
\displaystyle \big(\partial_t - \partial_{zz} \big) W^{(2,A)}(z, t) = W^{(0)}(z, t) \\[1ex]
\displaystyle \partial_z W^{(2,A)}(z, t) \Big|_{z=0}=0 \\[1ex]
\displaystyle W^{(2,A)}(z, 0) = 0 
\end{dcases}
\label{IBVP_W2A}
\end{equation}

\subsubsection{Reduced IBVP for $W^{(2,B)}(z, t)$ in $T^{(2,B)}$}
\begin{equation}
\begin{dcases}
\displaystyle \big(\partial_t - \partial_{zz} \big) W^{(2,B)}(z, t) = 
\frac{z^2}{\lambda} e^{-z/\lambda }, \quad 
\lambda \equiv \cos\theta_2 \\[1ex]
\displaystyle \partial_z W^{(2,B)}(z, t) \Big|_{z=0}=0 \\[1ex]
\displaystyle W^{(2,B)}(z, 0) = 0 
\end{dcases}
\label{IBVP_W2B}
\end{equation}

\subsection{Closed Form Solutions of Asymptotic Terms}
We solve for $W^{(0)}$, $W^{(1)}$, $W^{(2,A)}$ and $W^{(2,B)}$
from their governing IBVPs obtained above. 
\subsubsection{Analytical Expression of $W^{(0)}(z, t)$}
Previously, we solved model \eqref{IBVP_1} in the situation of i) 
perpendicular incident beam and ii) negligible lateral heat conduction, 
which corresponds to IBVP \eqref{IBVP_W0} with $\lambda = 1$ 
\cite{WBZ_2020C}\cite{Jaime-Yepez_2024}. 
The analytical solution of \eqref{IBVP_W0} for $\lambda = 1$, 
$U^{(0)}(z, t) \equiv W^{(0)}(z, t; \lambda)\big|_{\lambda=1}$, 
is a parameter free function and will be used repeatedly in this study. 
\begin{equation}
\begin{aligned}
& U^{(0)}(z, t) \equiv W^{(0)}(z, t; \lambda)\big|_{\lambda=1} = -e^{-z}+\frac{e^{-z+t}}{2} 
\text{erfc}(\frac{-z+2t}{\sqrt{4t} }) \\
& \qquad + \frac{e^{z+t}}{2} \text{erfc}(\frac{z+2t}{\sqrt{4t} }) 
-z \, \text{erfc}(\frac{z}{\sqrt{4 t} }) 
+ \frac{2\sqrt{t}}{\sqrt{\pi }} e^{\frac{-z^2}{4t}} 
\end{aligned} 
\label{U0_exp}  
\end{equation}
For general $\lambda = \cos \theta_2 $, we use a linear scaling to transform \eqref{IBVP_W0} 
to the case of $\lambda = 1$. Let 
\[ \tilde{z} \equiv \frac{z}{\lambda}, \qquad \tilde{t} \equiv \frac{t}{\lambda^2}, \qquad 
\tilde{W}^{(0)}(\tilde{z}, \tilde{t}) \equiv \frac{1}{\lambda}W^{(0)}(z, t; \lambda) \] 
It is straightforward to verify that $ \tilde{W}^{(0)}(\tilde{z}, \tilde{t})$ satisfies 
IBVP \eqref{IBVP_W0} in variables $(\tilde{z}, \tilde{t})$ with $\lambda = 1$. 
It follows that $\tilde{W}^{(0)}(\tilde{z}, \tilde{t})
= U^{(0)}(\tilde{z}, \tilde{t}) $. 
Thus, the solution of \eqref{IBVP_W0} for general $\lambda $ is  
\begin{equation}
\boxed{\quad W^{(0)}(z, t; \lambda) = \lambda U^{(0)}(\tilde{z}, \tilde{t})
\Big|_{(\tilde{z}, \tilde{t})=(\frac{z}{\lambda}, \frac{t}{\lambda^2})} \quad}
\label{W0_exp} 
\end{equation}

\subsubsection{Analytical Expression of $W^{(1)}(z, t)$}
To solve $W^{(1)}(z, t)$ from \eqref{IBVP_W1}, we start with \eqref{IBVP_W0}. 
On both sides of \eqref{IBVP_W0}, 
we multiply by $\lambda$, differentiate with respect to $\lambda$, and then 
multiply by $\lambda$ to obtain 
\[ \displaystyle \big(\partial_t - \partial_{zz} \big) 
\big[\lambda \partial_\lambda \big( \lambda W^{(0)}(z, t; \lambda) \big) \big] = 
\lambda \partial_\lambda \big(\lambda \frac{1}{\lambda} e^{-z/\lambda} \big) 
= \frac{z}{\lambda} e^{-z/\lambda } \] 
Thus, $\lambda \partial_\lambda \big( \lambda W^{(0)}(z, t; \lambda) \big) $ 
is the solution $W^{(1)}(z, t; \lambda)$ of \eqref{IBVP_W1}. 
We evaluate $ \lambda \partial_\lambda \big( \lambda W^{(0)}(z, t; \lambda) \big)$
using $W^{(0)}(z, t; \lambda) = \lambda U^{(0)}(\tilde{z}, \tilde{t})$ from \eqref{W0_exp}. 
\[ W^{(1)}(z, t; \lambda) = \lambda \partial_\lambda \big( \lambda^2 
U^{(0)}(\tilde{z}, \tilde{t}) \big), 
\qquad (\tilde{z}, \tilde{t})=(\frac{z}{\lambda}, \frac{t}{\lambda^2}) \]
\begin{equation}
\boxed{ \quad W^{(1)}(z, t; \lambda)= \lambda^2 \Big( 2 U^{(0)}(\tilde{z}, \tilde{t}) 
-\tilde{z} U^{(0)}_{\tilde{z}}(\tilde{z}, \tilde{t})
-2\tilde{t} U^{(0)}_{\tilde{t}}(\tilde{z}, \tilde{t})
\Big) \Big|_{(\tilde{z}, \tilde{t})=(\frac{z}{\lambda}, \frac{t}{\lambda^2})}\quad }
\label{W1_exp}
\end{equation}
where $U^{(0)}(z, t)$ is given in \eqref{U0_exp} and its first derivatives are 
\begin{equation}
\begin{dcases}
U^{(0)}_z(z, t) = e^{-z}-\frac{e^{-z+t}}{2} \text{erfc}(\frac{-z+2t}{\sqrt{4t}})
+\frac{e^{z+t}}{2} \text{erfc}(\frac{z+2t}{\sqrt{4t}})-\text{erfc}(\frac{z}{\sqrt{4 t}}) \\[2ex]
U^{(0)}_t(z, t) = \frac{e^{-z+t}}{2} \text{erfc}(\frac{-z+2t}{\sqrt{4t}})
+\frac{e^{z+t}}{2} \text{erfc}(\frac{z+2t}{\sqrt{4t}}) 
\end{dcases}
\label{U0_d1}
\end{equation}

\subsubsection{Analytical Expression of $W^{(2,A)}(z, t)$}
Previously, we solved a two-term asymptotic solution incorporating the effect of 
lateral heat conduction for a perpendicular incident beam. 
The governing IBVP for the $O(\varepsilon^2)$ term in that case 
corresponds to IBVP \eqref{IBVP_W2A} with $\lambda = 1$ \cite{Jaime-Yepez_2024}. 
The expression of the $O(\varepsilon^2)$ term in that case, 
$U^{(2)}(z, t) \equiv W^{(2,A)}(z, t; \lambda)\big|_{\lambda=1}$, 
is a parameter free function. 
\begin{equation}
\begin{aligned}
& U^{(2)}(z, t) \equiv W^{(2,A)}(z, t; \lambda)\big|_{\lambda=1} = 
e^{-z} + (t-1) \frac{e^{-z+t}}{2} \text{erfc}(\frac{-z+2t}{\sqrt{4t }}) 
 \\[2ex]
& \qquad + (t-1) \frac{e^{z+t} }{2} \text{erfc}(\frac{z+2t}{\sqrt{4t}}) 
+\frac{z}{6} (6+z^2 ) \text{erfc}(\frac{z}{\sqrt{4t }}) 
+ \frac{\sqrt{t}}{3\sqrt{\pi }} (2t-6-z^2) e^{\frac{-z^2}{4t}}  
\end{aligned}
\label{U2_exp}
\end{equation}
For general $\lambda > 0$, the source term in \eqref{IBVP_W2A} is 
\[ \text{source}(\lambda) = W^{(0)}(z, t) = \lambda U^{(0)}(\tilde{z}, \tilde{t}), 
\qquad (\tilde{z}, \tilde{t})=(\frac{z}{\lambda}, \frac{t}{\lambda^2}) \] 
We use a linear scaling to transform \eqref{IBVP_W2A} 
to the case of $\lambda = 1$. Let 
\[ \tilde{z} \equiv \frac{z}{\lambda}, \qquad \tilde{t} \equiv \frac{t}{\lambda^2}, \qquad 
\tilde{W}^{(2,A)}(\tilde{z}, \tilde{t}) \equiv \frac{1}{\lambda^3}W^{(2,A)}(z, t; \lambda) \] 
It is straightforward to verify that $ \tilde{W}^{(2,A)}(\tilde{z}, \tilde{t})$ satisfies 
IBVP \eqref{IBVP_W2A} in variables $(\tilde{z}, \tilde{t})$ with $\lambda = 1$. 
It follows that $\tilde{W}^{(2,A)}(\tilde{z}, \tilde{t})
= U^{(2)}(\tilde{z}, \tilde{t}) $. Thus, the solution for general $\lambda $ is 
\begin{equation}
\boxed{\quad W^{(2,A)}(z, t; \lambda) = \lambda^3 U^{(2)}(\tilde{z}, \tilde{t})
\Big|_{(\tilde{z}, \tilde{t})=(\frac{z}{\lambda}, \frac{t}{\lambda^2})} \quad} 
\label{W2A_exp} 
\end{equation}

\subsubsection{Analytical Expression of $W^{(2,B)}(z, t)$}
To solve $W^{(2,B)}(z, t)$ from \eqref{IBVP_W2B}, we again start with \eqref{IBVP_W0}. 
On both sides of \eqref{IBVP_W0}, 
we repeat multiplying by a power of $\lambda$ and differentiating with respect to $\lambda$
 to obtain 
\[ \displaystyle \big(\partial_t - \partial_{zz} \big) 
\Big( \lambda \partial_\lambda \big[\lambda^2 \partial_\lambda \big( \lambda W^{(0)}(z, t; \lambda) 
\big) \big] \Big) 
=  \lambda \partial_\lambda \big[
\lambda^2 \partial_\lambda \big(\lambda \frac{1}{\lambda} e^{-z/\lambda} \big) \big]
= \frac{z^2 }{\lambda} e^{-z/\lambda } \] 
Thus, $\lambda \partial_\lambda \big[\lambda^2 \partial_\lambda 
\big( \lambda W^{(0)}(z, t; \lambda) \big) \big] $ 
is the solution $W^{(2,B)}(z, t; \lambda)$ of \eqref{IBVP_W2B}. 
We evaluate \\ $\lambda \partial_\lambda \big[\lambda^2 \partial_\lambda 
\big( \lambda W^{(0)}(z, t; \lambda) \big) \big] $
using $W^{(0)}(z, t; \lambda) = \lambda U^{(0)}(\tilde{z}, \tilde{t})$ from \eqref{W0_exp}. 
\[ W^{(2,B)}(z, t; \lambda) = 
\lambda \partial_\lambda \big[\lambda^2 \partial_\lambda 
\big( \lambda^2 U^{(0)}(\tilde{z}, \tilde{t}) \big) \big], 
\qquad (\tilde{z}, \tilde{t})=(\frac{z}{\lambda}, \frac{t}{\lambda^2}) \]
\begin{equation}
\boxed{\quad \begin{aligned}
& W^{(2,B)}(z, t; \lambda) = \lambda^3 \Big( 6 U^{(0)}(\tilde{z},\tilde{t}) -4 \tilde{z} U^{(0)}_{\tilde{z}}(\tilde{z},\tilde{t}) -6 \tilde{t} U^{(0)}_{\tilde{t}}(\tilde{z},\tilde{t}) \\[1ex]
& \qquad \qquad + \tilde{z}^2 U^{(0)}_{\tilde{z}\tilde{z}}(\tilde{z},\tilde{t}) 
+ 4\tilde{z} \tilde{t}\, U^{(0)}_{\tilde{z}\tilde{t}}(\tilde{z},\tilde{t}) 
+4\tilde{t}^2 U^{(0)}_{\tilde{t}\tilde{t}}(\tilde{z},\tilde{t}) \Big)
\Big|_{(\tilde{z},\tilde{t}) = (\frac{z}{\lambda},\frac{t}{\lambda^2})} 
\end{aligned} \quad} \label{W2B_exp}
\end{equation}
where $U^{(0)}(z, t)$ is given in \eqref{U0_exp}, the first derivatives of $U^{(0)}(z, t)$ 
are given in\eqref{U0_d1}, and the second derivatives of $U^{(0)}(z, t)$ are 
\begin{equation}
\begin{dcases}
U^{(0)}_{zz}(z, t) = -e^{-z}+\frac{e^{-z+t}}{2} \text{erfc}(\frac{-z+2t}{\sqrt{4t}})
+\frac{e^{z+t}}{2} \text{erfc}(\frac{z+2t}{\sqrt{4t}})\\[2ex]
U^{(0)}_{zt}(z, t) = -\frac{e^{-z+t}}{2} \text{erfc}(\frac{-z+2t}{\sqrt{4t}})
+\frac{e^{z+t}}{2} \text{erfc}(\frac{z+2t}{\sqrt{4t}}) \\[2ex]
U^{(0)}_{tt}(z, t) = \frac{e^{-z+t}}{2} \text{erfc}(\frac{-z+2t}{\sqrt{4t}})
+\frac{e^{z+t}}{2} \text{erfc}(\frac{z+2t}{\sqrt{4t}})- \frac{1}{\sqrt{\pi t}} e^{\frac{-z^2}{4t}}
\end{dcases}
\label{U0_d2}
\end{equation}
%

\section{Effects of Incident Angle and Lateral Heat Conduction}
We dissect the contributions from lateral heat conduction and from the beam 
incident angle, and compare the magnitudes of these two contributions in the full
temperature solution. 
In this process, we also evaluate the performance of various asymptotic solutions 
from only the leading term to the one containing all $O(\varepsilon^2)$ terms. 

As we discussed in Section \ref{asy_formu}, 
the lateral heat conduction takes effect only in the term 
$\varepsilon^2 (\frac{\partial^2 T}{\partial x^2}+\frac{\partial^2 T}{\partial y^2})$
in the governing model \eqref{IBVP_1}. 
The effect of lateral heat conduction is regulated by
the depth-to-lateral scale ratio $\varepsilon$. 
In contrast, the effects of incident angle are spread over many places. 
Only one of the effects, the depth dependent lateral shift in the heat source 
$f(x+\varepsilon z \tan\theta_2, y)$ is regulated by $\varepsilon$. 
The rest are independent of $\varepsilon$. 
For example, all of $P_d^{(a)} \equiv (\alpha \cos\theta_1  P_d^{(i)})$, 
$f(x, y) \equiv q(x \cos\theta_1, y)$ and $\lambda = \cos\theta_2$ 
vary with the incident angle but their effect are independent of $\varepsilon$. 
That is, in the hypothetical situation of $\varepsilon = 0$, these effects are still present. 
Our asymptotic formulation is based on a small $\varepsilon $, in which 
the higher order terms represent the effects that are modulated by $\varepsilon $. 
For that reason, when we discuss the effect of incident angle in constructing 
an asymptotic approximation, we focus on 
the depth dependent lateral shift in the heat source $f(x+\varepsilon z \tan\theta_2, y)$. 

\subsection{Extended Formulation for Dissecting the Effects of Lateral conduction and Incident Angle}
In our asymptotic formulation, when constructing the $O(\varepsilon^2)$ terms,
we used parts: $T^{(2,A)}$ for the $O(\varepsilon^2)$ effect of lateral conduction
and $T^{(2,B)}$ for the $O(\varepsilon^2)$ effect of incident angle.  
Physically, $\varepsilon$ in 
$\varepsilon^2 (\frac{\partial^2 T}{\partial x^2}+\frac{\partial^2 T}{\partial y^2})$
and the one in $f(x+\varepsilon z \tan\theta_2, y)$ 
are the same parameter: the depth-to-lateral scale ratio. 
To distinguish the effects from lateral conduction and from the incident angle, 
we work with a mathematically extended formulation in which these two 
$\varepsilon$'s are distinguished. We use $\varepsilon_1$ for lateral conduction
and $\varepsilon_2$ for modulating the depth dependent lateral shift. 
The extended formulation is 
\begin{equation}
\begin{dcases}
\displaystyle \frac{\partial T}{\partial t} 
= \varepsilon_1^2 \Big(\frac{\partial^2T}{\partial x^2}
+\frac{\partial^2T}{\partial y^2}\Big)+\frac{\partial^2T}{\partial z^2}
+P_d^{(a)} f(x+\varepsilon_2 z \tan\theta_2, y) 
\frac{1}{\lambda} e^{-z/\lambda}, 
\quad \lambda \equiv \cos\theta_2 \\[2ex]
\displaystyle \frac{\partial T(x, y, z, t)}{\partial z} \bigg|_{z=0}=0 \\[1ex]
\displaystyle T(x, y, z, 0) = 0 
\end{dcases}
\label{IBVP_1rev}
\end{equation}
The extension to \eqref{IBVP_1rev} is purely mathematical. It allows us 
to turn on/off separately the lateral conduction and the effect of incident angle. 
In the physical model \eqref{IBVP_1}, these two effects are connected by the same 
physical $\varepsilon$. 
The asymptotic formulation corresponding to the extended formulation \eqref{IBVP_1rev} is
\begin{align}
& T(x, y, z, t) = T^{(0)}(x, y, z, t) + \varepsilon_2\, T^{(1)}(x, y, z, t) \nonumber \\[1ex]
& \qquad \qquad + \underbrace{\varepsilon_1^2\, T^{(2,A)}(x, y, z, t)}_{\text{lateral conduction}} 
+ \varepsilon_2^2\, T^{(2,B)}(x, y, z, t) + \cdots 
\label{asymp_form_rev}
\end{align}
All asymptotic terms have separable dependence on $(x, y)$ and on $(z, t)$. 
They were derived in Section \ref{asy_sol} and are summarized below. 
\begin{align}
& T^{(0)}(x, y, z, t) = \Big(P_d^{(a)} f(x, y)\Big) W^{(0)}(z, t)  \label{T0_W0} \\[2ex]
& T^{(1)}(x, y, z, t) = \Big(P_d^{(a)} f_x(x, y) \tan{\theta_2} \Big) W^{(1)}(z, t) 
 \label{T1_W1} \\[2ex]
& T^{(2,A)}(x, y, z, t) = \Big(P_d^{(a)} \nabla^2_{(x, y)} f(x, y) \Big) 
W^{(2,A)}(z, t) \label{T2A_W2A} \\[2ex] 
& T^{(2,B)}(x, y, z, t) = \Big(P_d^{(a)} \frac{1}{2}f_{xx}(x, y) \tan^2{\theta_2} \Big) 
W^{(2,B)}(z, t) \label{T2B_W2B}
\end{align}
where $W^{(0)}(z, t)$, $W^{(1)}(z, t)$, $W^{(2,A)}(z, t)$ and $W^{(2,B)}(z, t)$ 
are given respectively in \eqref{W0_exp}, \eqref{W1_exp}, \eqref{W2A_exp}
and \eqref{W2B_exp}. 

\subsection{Errors in Various Asymptotic Solutions}
The extended asymptotic formulation \eqref{asymp_form_rev} with $\varepsilon_1$
and $\varepsilon_2$ gives several asymptotic approximations, 
from only the leading term to the one containing both $O(\varepsilon_1^2)$
and $O(\varepsilon_2^2)$ terms. 
We calculate the errors in these asymptotic solutions by comparing them 
to an effectively ``exact'' solution of the full model \eqref{IBVP_1rev}. 

So far, there is no closed form analytic solution of IBVP \eqref{IBVP_1rev} 
that can be computed efficiently. 
Instead we find a very accurate solution numerically. 
We use the central finite difference method to solve IBVP \eqref{IBVP_1rev} 
on a fine numerical grid over a region in the 3D skin. 
The numerical grid is fine enough to reduce the discretization error. 
The computational region is large enough in both the depth and lateral directions so that 
the error induced by cutting off at large $z$ and/or large $(x, y)$ is 
smaller than the discretization error. 
The simulations were implemented in Python on clusters with GPUs. 
The utilization of GPUs serves two purposes: i) to speed up the finite difference simulation
on a fine spatial grid over a large 3D region, and ii) to facilitate the large memory 
allocation required for accommodating both a fine grid and a large 3D computational region. 

The extended formulation \eqref{IBVP_1rev} allows us to mathematically
set  $\varepsilon_1$ (lateral conduction) and $\varepsilon_2$ 
(modulating the effect of incident angle) separately. 
Below we carry out simulation analysis on \eqref{IBVP_1rev} for several hypothetical 
combinations of $(\varepsilon_1, \varepsilon_2)$ motivated by a moderate value of  
$\varepsilon = 0.1$. The objective of testing various combinations of 
$(\varepsilon_1, \varepsilon_2)$ is to examine separately the contributions from 
lateral conduction and from the incident angle, and to check the accuracy of asymptotic solutions. 
In the simulations, we use $t_\text{final} = 4$ and the beam intrinsic relative distribution
of power density over its perpendicular cross-section
\[ q(x, y) = \exp(\frac{-(x^2+y^2)}{2}) \]
which is in non-dimensionalized spatial variables $(x, y)$. 
The corresponding relative distribution of power projected onto the skin surface is given by 
\[f(x, y) = q(\cos\theta_1 x, y) = \exp(\frac{-(\cos^2\theta_1 x^2+y^2)}{2}) \]

\subsubsection{Case 1: no lateral conduction, zero incident angle}
We start with the case of $\varepsilon_1 = 0 $ and 
$\theta_1 = \theta_2 = 0$ ($\varepsilon_2$ does not matter in this case since $\tan\theta_2 = 0$). 
Case 1 is the situation where the incident beam is perpendicular to skin and 
the lateral heat conduction is neglected. In case 1, the leading term 
$T^{(0)}(x, y, z, t)$ given in \eqref{T0_W0} is the exact solution of \eqref{IBVP_1rev}. 
The objective of solving this special case to gauge the accuracy of the 3D numerical solver 
and to test the reliability of numerical error estimation. 
To estimate the error of numerical solution when no exact solution is available, 
we carry out simulations  on a grid and again on the grid refined by a factor of 2. 
Let 
\begin{itemize}
\item $T_\text{exact} $ be the exact solution;  
\item $T_\text{coarse} $ the numerical solution on the coarse grid; 
\item $E_\text{coarse} \equiv T_\text{coarse}- T_\text{exact}$ the error in $T_\text{coarse} $; 
\item $T_\text{fine} $ the numerical solution on the refined grid; 
\item $E_\text{fine} \equiv T_\text{fine}- T_\text{exact}$ the error in $T_\text{fine} $.
\end{itemize}
Since the central finite difference has the second order accuracy, we expect 
\begin{align}
& E_\text{coarse} \approx 4E_\text{fine}
\quad \Longrightarrow \quad (T_\text{coarse}- T_\text{exact}) 
\approx 4(T_\text{fine}- T_\text{exact}) \nonumber \\[1ex]
& \boxed{\quad \big| T_\text{fine}- T_\text{exact}\big| \approx \frac{1}{3} 
\Big| T_\text{coarse} - T_\text{fine} \Big| \quad} 
\label{num_err_est}
\end{align}
Formula \eqref{num_err_est} provides a practical numerical tool for estimating the error 
$\big| E_\text{fine}\big| \equiv \big| T_\text{fine}- T_\text{exact}\big| $ 
when the exact solution $T_\text{exact}$ is unknown. 
\begin{figure}[!h]
\vskip 0.4cm
\begin{center}
\psfig{figure=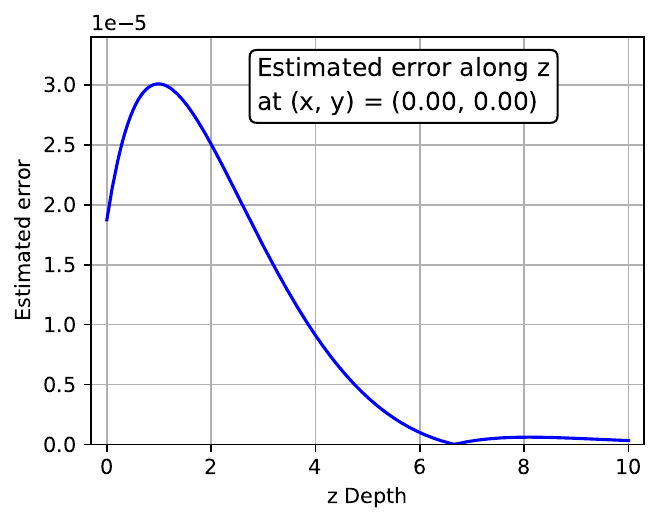, width=3.0in}\;\; \psfig{figure=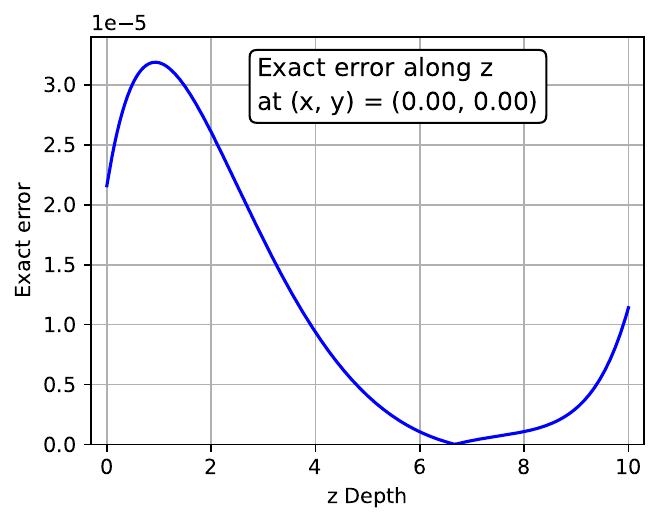, width=3.0in}
\end{center}
\vskip -0.8cm
\caption{Case 1: no lateral conduction ($\varepsilon_1 = 0$), zero incident angle 
($\theta_1 = \theta_2 = 0$).
Left: estimated error $\frac{1}{3}\big| T_\text{coarse}-T_\text{fine}\big| $. 
Right: exact error $\big| T_\text{fine}-T_\text{exact}\big| $. } 
\label{fig_03}
\end{figure}
Figure \ref{fig_03} compares the exact error in numerical solution $T_\text{fine}$ 
(calculated using $T_\text{exact}$) and the estimated error in $T_\text{fine}$
(calculated using $T_\text{fine}$ and $T_\text{coarse}$ as described
in \eqref{num_err_est}). 
The results of Figure \ref{fig_03} confirm the reliability of numerical error estimation. 
In the subsequent cases, we shall use the numerical error estimation to 
establish the accuracy of the numerical solution and use the numerical solution 
to evaluate the errors of asymptotic solutions. 

\subsubsection{Case 2: no lateral conduction, incident angle = 30 degree}
We examine the case of $\varepsilon_1 = 0$, 
$\theta_1 = 30^\circ$, $\theta_2 = 20^\circ$, and $\varepsilon_2 = 0.1$. 
Here the effect of lateral conduction is artificially removed by setting $\varepsilon_1 = 0$. 
The objective is to investigate the effect of a non-trivial incident angle
in the absence of lateral heat conduction.

\begin{figure}[!h]
\vskip 0.4cm
\begin{center}
\psfig{figure=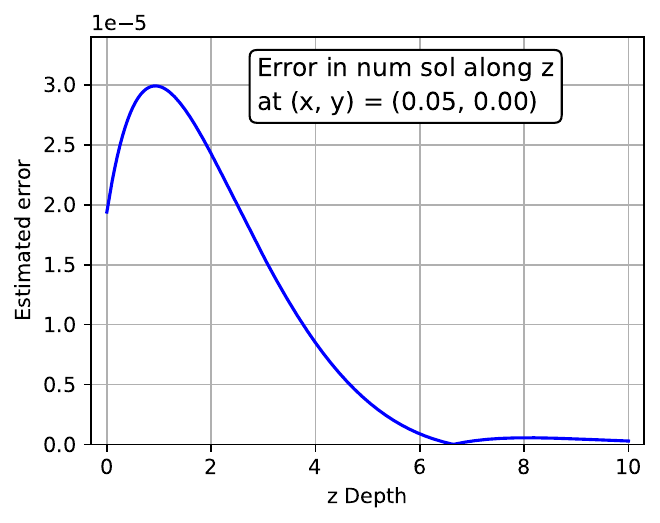, width=3.0in}\;\; \psfig{figure=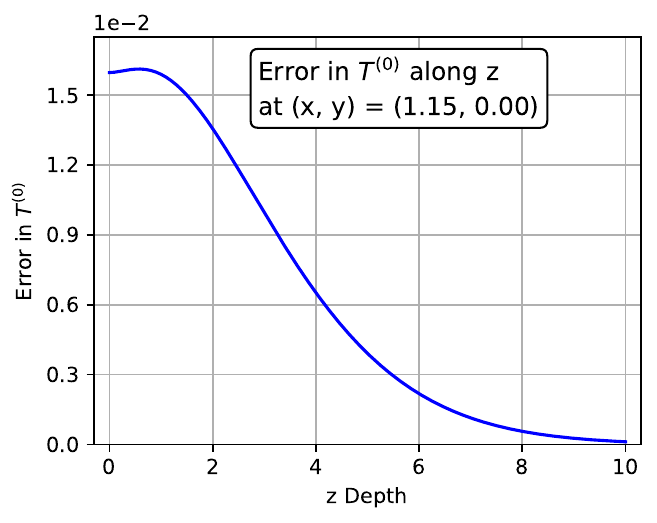, width=3.0in} \\
\psfig{figure=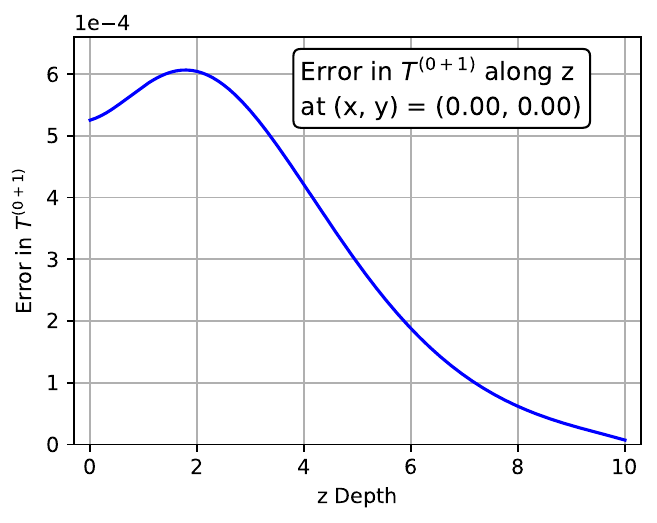, width=3.0in}\;\; \psfig{figure=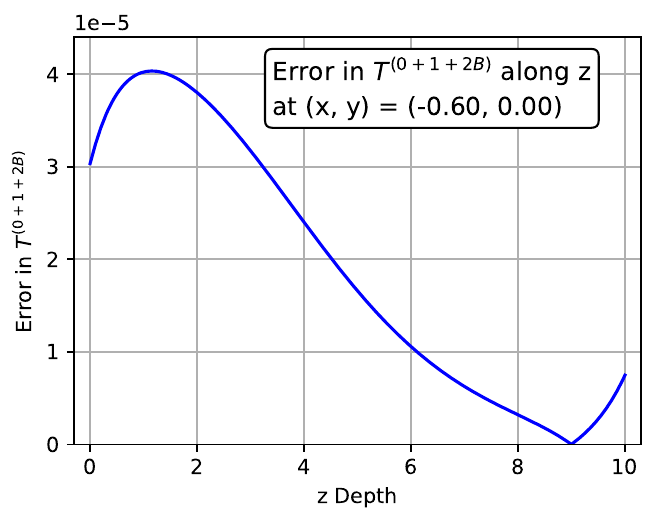, width=3.0in}
\end{center}
\vskip -0.8cm
\caption{Case 2: no lateral conduction ($\varepsilon_1 =0$), 
incident angle ($\theta_1 = 30^\circ$, $\theta_2 = 20^\circ$) modulated by $\varepsilon_2 =0.1$.
Top left: estimated error in numerical solution. 
Top right: error in $T^{(0)}$. 
Bottom left: error in $T^{(0+1)}$. 
Bottom right: error in $T^{(0+1+2B)}$. } 
\label{fig_04}
\end{figure}
Figure \ref{fig_04} displays the results of case 2. The top left panel establishes 
the accuracy of the numerical solution. The other three panels use this numerical solution
to evaluate the errors in three asymptotic solutions, which are concisely denoted by 
\begin{align*}
& T^{(0+1)} \equiv T^{(0)}+\varepsilon_2\, T^{(1)} \\[1ex]
& T^{(0+1+2B)} \equiv T^{(0)}+\varepsilon_2\, T^{(1)}+\varepsilon_2^2\, T^{(2,B)}
\end{align*}
The results of Figure \ref{fig_04} indicates that at a moderate $\varepsilon = 0.1$
and in the absence of lateral conduction, including each additional term in the asymptotic 
solution significantly improve the approximation accuracy. 

\subsubsection{Case 3: positive lateral heat conduction, zero incident angle}
\begin{figure}[!h]
\vskip 0.4cm
\begin{center}
\psfig{figure=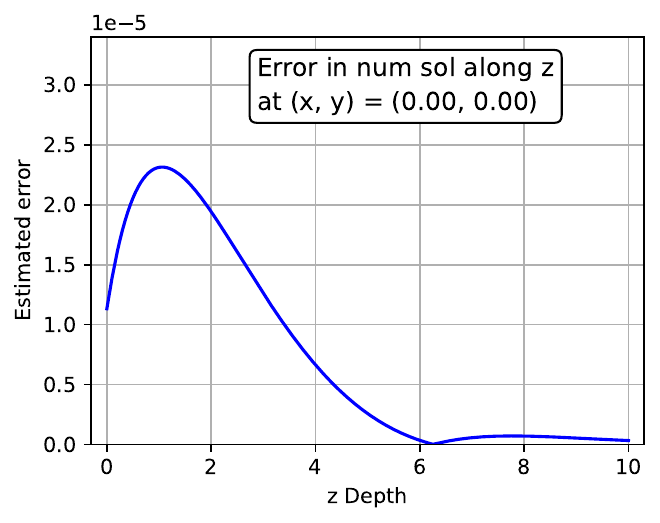, width=3.0in}\;\; \psfig{figure=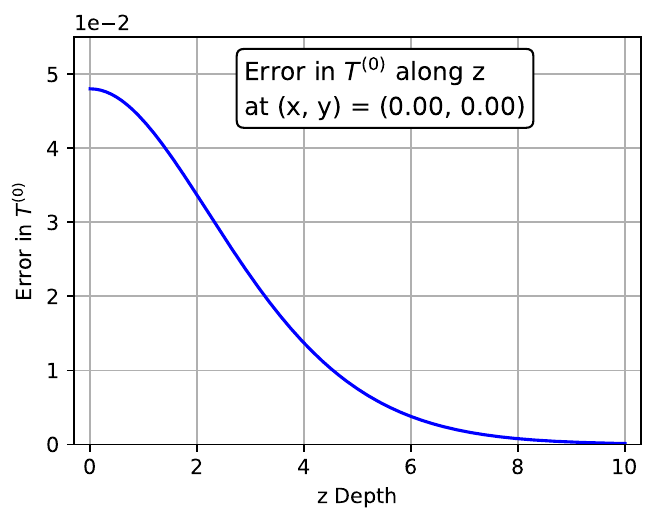, width=3.0in} \\
\psfig{figure=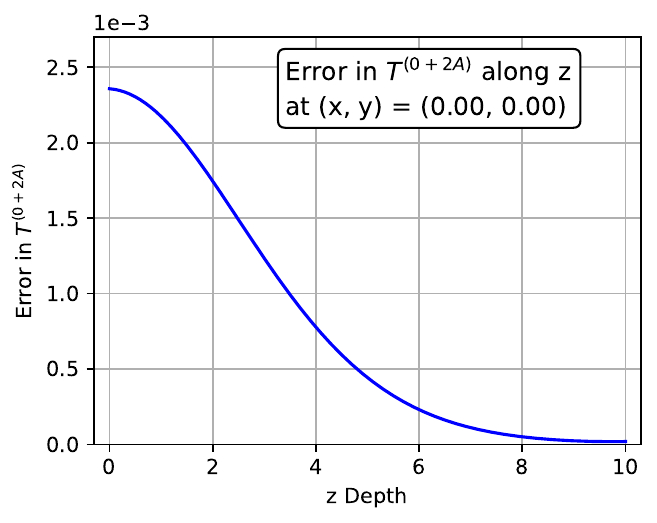, width=3.0in}
\end{center}
\vskip -0.8cm
\caption{Case 3: positive lateral conduction ($\varepsilon_1 =0.1$), 
zero incident angle ($\theta_1 = \theta_2 = 0$).
Top left: estimated error in numerical solution. Top right: error in $T^{(0)}$. 
Bottom: error in $T^{(0+2A)}$. 
Here $T^{(1)} = T^{(2,B)} = 0$ because $\tan\theta_2 = 0$. } 
\label{fig_05}
\end{figure}

We examine the case of $\varepsilon_1 = 0.1$, $\theta_1 = \theta_2 =0$ ($\varepsilon_2$ 
does not matter in this case since $\tan \theta_2 = 0$). 
Case 3 allows us to focus on the effect of lateral heat conduction in the absence of 
incident angle effect. 
Figure \ref{fig_05} gives the results of case 3. The top left panel establishes 
the accuracy of the numerical solution. The other two panels use this numerical solution
to evaluate the errors in two asymptotic solutions.  

Note that because $\tan \theta_2 = 0$, we have $T^{(1)} = T^{(2,B)} = 0$ in case 3. 
The top right panel of Figure \ref{fig_05} shows the error in the leading term $T^{(0)}$
when the true solution contains $\varepsilon^2 T^{(2,A)}$ 
but not $\varepsilon T^{(1)}$.  
In comparison, the top right panel of Figure \ref{fig_04} plots the error in $T^{(0)}$
when the true solution contains $\varepsilon T^{(1)}$ but not $\varepsilon^2 T^{(2,A)}$. 
Mathematically, $O(\varepsilon^2) \ll O(\varepsilon)$ as $\varepsilon \rightarrow 0$. 
We expect that the error of $T^{(0)}$ in case 3 (Figure \ref{fig_05}) is below 
the error of $T^{(0)}$ in case 2 (Figure \ref{fig_04}) as $\varepsilon \rightarrow 0$. 
However, at moderate $\varepsilon=0.1$, 
the error of $T^{(0)}$ is significantly higher in case 3 (Figure \ref{fig_05}) than 
in case 2 (Figure \ref{fig_04}). This observation indicates that at a moderate $\varepsilon$, 
the $O(\varepsilon^2)$ contribution from the lateral conduction $\varepsilon^2 T^{(2,A)}$ 
may be larger than the $O(\varepsilon)$ effect of the incident angle $\varepsilon T^{(1)}$. 
Thus, as a moderate $\varepsilon$, adding only $T^{(1)}$ without $T^{(2,A)}$
in the asymptotic solution may not give us a meaningful improvement in accuracy at all. 

\subsubsection{Case 4: Positive lateral heat conduction, incident angle = 30 degree}
\begin{figure}[!h]
\vskip 0.4cm
\begin{center}
\psfig{figure=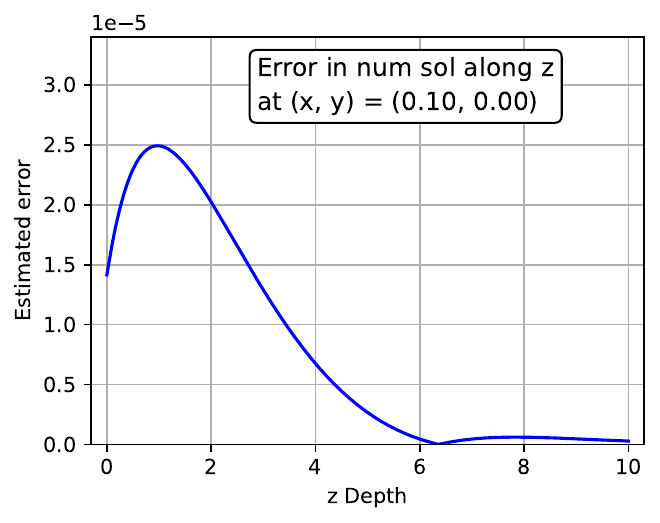, width=3.0in}\;\; \psfig{figure=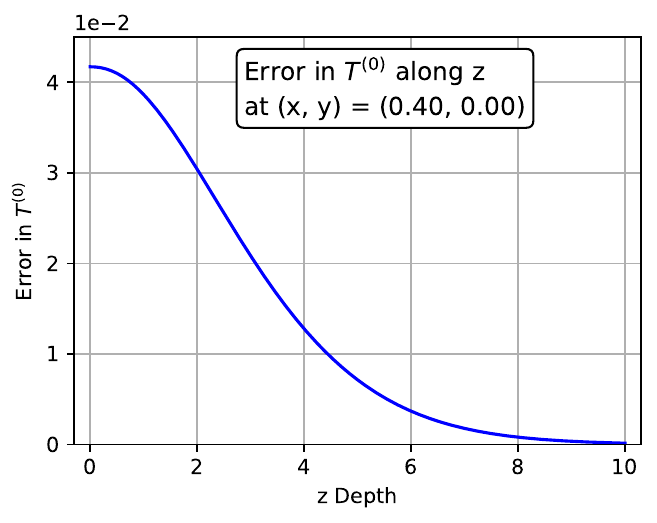, width=3.0in} \\
\psfig{figure=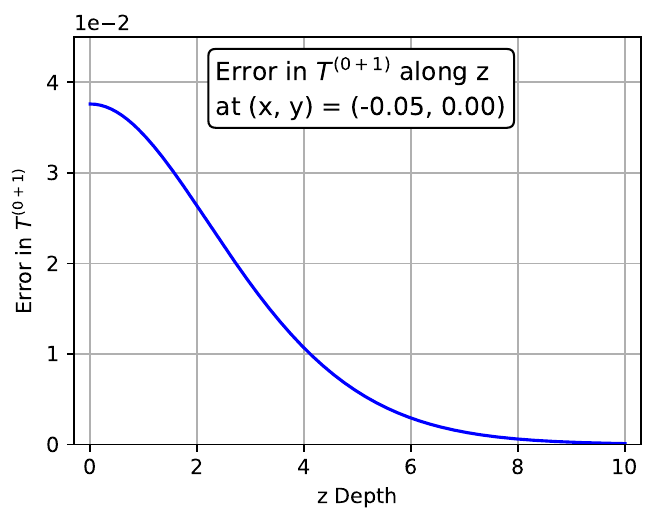, width=3.0in}\;\; \psfig{figure=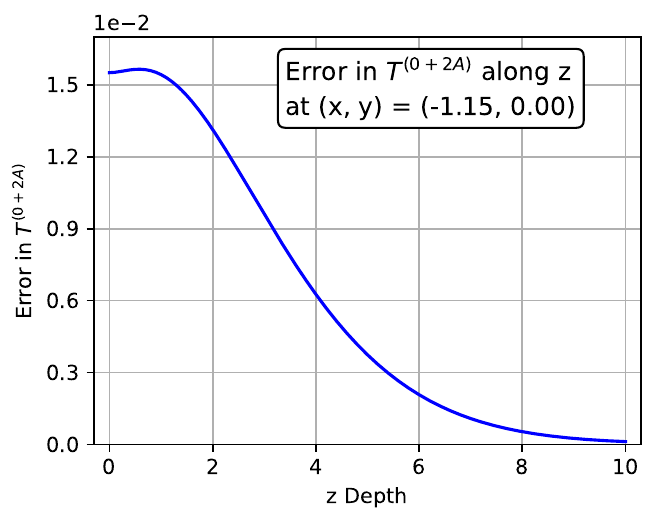, width=3.0in} \\
\psfig{figure=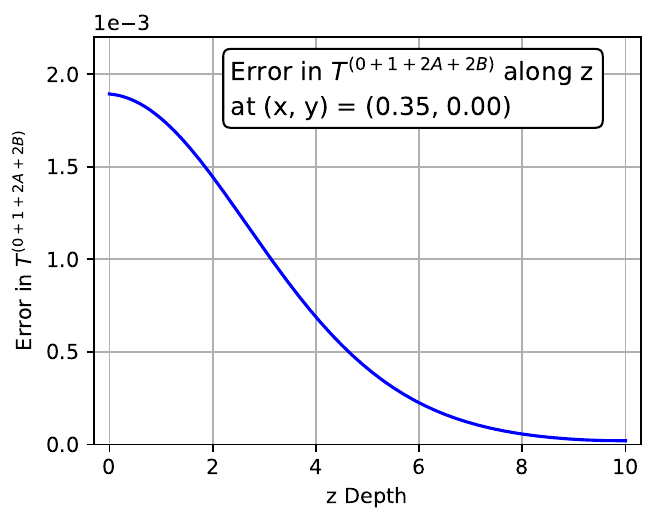, width=3.0in}
\end{center}
\vskip -0.8cm
\caption{Case 4: $\varepsilon_1 = \varepsilon_2=0.1$, 
$\theta_1 = 30^\circ$, $\theta_2 = 20^\circ$. 
Top left: estimated error in numerical solution. Top right: error in $T^{(0)}$. 
Middle left: error in $T^{(0+1)}$. Middle right: error in $T^{(0+2A)}$. 
Bottom: error in $T^{(0+1+2A+2B)}$. } 
\label{fig_06}
\end{figure}

We examine the case of $\varepsilon_1 = \varepsilon_2 = 0.1$, $\theta_1 = 30^\circ$,
$ \theta_2 =20^\circ$. 
This is the physical situation of a non-trivial incident angle in the presence of lateral heat conduction.
In this case, the hypothetical $\varepsilon_1$ and $\varepsilon_2$ are united with 
the physical $\varepsilon$. 
The true solution contains effects of both the lateral heat conduction 
and the incident angle. 
Figure \ref{fig_06} presents the results of case 4. 
The top left panel establishes the accuracy of the numerical solution. 
The other four panels use this numerical solution to evaluate the errors 
in four asymptotic solutions. 
Comparing the top right and the middle left panels, we see that adding the 
$\varepsilon T^{(1)}$ term in the asymptotic solution barely improves 
accuracy at all. In contrast, adding the $\varepsilon^2 T^{(2,A)}$ term without 
adding $\varepsilon T^{(1)}$ in the asymptotic solution noticeably reduces the error from 
$4.2\times 10^{-2}$ (top right) to $1.6\times 10^{-2}$ (middle right). Note that 
$T^{(0+2A)} \equiv T^{(0)}+\varepsilon^2 T^{(2,A)}$ in the middle right panel
is not a meaningful approximation in the limit of $\varepsilon \rightarrow 0$. 
It is an ad hoc test to illustrate that at $\varepsilon = 0.1$, 
the $O(\varepsilon^2)$ contribution from lateral conduction outweighs the 
$O(\varepsilon)$ contribution from incident angle. 
The bottom panel shows the error in the asymptotic solution containing all terms up to 
and including both $O(\varepsilon^2)$ terms. The second order asymptotic solution 
$T^{(0+1+2A+2B)} \equiv T^{(0)}++\varepsilon T^{(1)}
+\varepsilon^2 T^{(2,A)}+\varepsilon^2 T^{(2,B)}$ demonstrates a fairly good accuracy 
even at the moderate $\varepsilon = 0.1$. 
%

\clearpage
\section{Concluding Remarks}
We study the problem of the 3D temperature evolution when a skin area is exposed 
to an electromagnetic beam. Previously we examined the situation of incident beam 
perpendicular to the skin surface. Here we consider the general case where the incident beam 
is at an arbitrary angle $\theta_1$ to the normal direction of skin surface, which is referred to 
as the incident angle. The perpendicular incident corresponds to $\theta_1 = 0$. 
We carry out an asymptotic analysis to capture the effect of the incident angle. 
The asymptotic formulation is not based on the incident angle being small. Rather it is based 
on that the depth-to-lateral scale ratio is small. This formulation is valid for arbitrary 
incident angle, not necessarily small. 

Skin is a lossy media. When an electromagnetic wave propagates inside skin, 
the electromagnetic power is attenuated/absorbed and becomes a heat source. 
The propagation distance until the surviving fraction of power dropping to $(1/e)$ 
is referred to as the penetration depth. 
For an electromagnetic wave in the frequency range of 30-300 GHz, the penetration 
depth into skin tissue is sub-millimeter. The size of beam cross-section is usually of the order of
a centimeter or larger. These two aspects lead to different length scales in depth and in lateral directions, 
resulting in a small depth-to-lateral scale ratio $\varepsilon$. 

When the incident beam is tilted away from the perpendicular normal direction of 
the skin surface, the incident angle 
induces several effects: i) it elongates the beam spot projected on the skin surface; 
ii) accordingly it dilutes the power density projected onto the skin surface and passing into skin; 
iii) the refracted angle tilts the propagation direction away from the perpendicular
normal direction inside skin; 
since the heat source comes from the attenuation/absorption of electromagnetic power, 
the refracted direction shifts the heat source in a lateral direction in a depth dependent fashion, 
like a shear motion; that is, at each depth, the heating region is shifted in a lateral direction 
by an amount that varies with the depth. 
Effect iii) above complicates the 3D temperature evolution in skin. 
As a result, with effect iii), there is no closed form exact solution to the full PDE model that can 
be computed efficiently. 

The small depth-to-lateral scale ratio $\varepsilon$ has two effects: a) it modulates  
the effect of lateral heat conduction, making it less prominent than that of depth heat conduction; 
b) it modulates the depth dependent lateral shifting of the heat source in the 
nondimensional formulation. 
Since the surviving electromagnetic power drops exponentially in the depth direction, 
the magnitude of heat source drops accordingly in the depth. The intensified heating 
near the skin surface creates a large temperature gradient in the depth. 
The depth heat conduction directly interacts with the large temperature gradient. 
In contrast, the temperature gradient in lateral directions is much smaller and 
accordingly the impact of lateral conduction on temperature evolution is 
less pronounced. 

We use two separate length scales respectively in depth and lateral directions 
to carry out nondimensionalization. 
In the nondimensional formulation \eqref{IBVP_1rev}, 
the lateral heat conduction appears as 
$\varepsilon^2 (\frac{\partial^2 T}{\partial x^2}+\frac{\partial^2 T}{\partial y^2})$
and the depth dependent lateral shift of the heat source appears as 
$f(x+\varepsilon z \tan\theta_2, y)$. 
We carry out an asymptotic analysis to capture these two effects. 
Specifically, we adopt the asymptotic formulation \eqref{asymp_form_rev}. 
We derive the governing initial boundary value problem (IBVP) for each 
asymptotic term, and we solve each IBVP to obtain a closed form analytical expression
that can be computed efficiently. 
The resulting asymptotic solution containing all terms up to and including $O(\varepsilon^2)$
has a fairly good accuracy even at a moderately large $\varepsilon = 0.1$. 
In contrast, numerical solution of the full PDE model involves advancing in time the temperature 
distribution on a fine grid over a large spatial region of $(x, y, z)$. The accurate numerical 
solution of the full 3D model is beyond the computing capability of a typical desktop computer. 
Therefore, the newly developed second order asymptotic solution serves as a 
fast and viable tool for computing the 3D temperature distribution in the situation where 
an skin area is exposed to an electromagnetic beam at an arbitrary incident angle. 

\clearpage
\noindent{\bf \large Acknowledgement and disclaimer}

\noindent \indent
The authors acknowledge the Joint Intermediate Force Capabilities Office of U.S. Department of Defense and the Naval Postgraduate School for supporting this work. The views expressed in this document are those of the authors and do not reflect the official policy or position of the Department of Defense or the U.S. Government.


\end{document}